%% file: tensorflow.tex
\documentclass[10pt,twocolumn]{article}

\usepackage{times}
\usepackage{fullpage}

\usepackage{authblk}
\usepackage{hyphenat}
\usepackage{times}
\usepackage{macros}
\usepackage{alltt}
\usepackage{xcolor}
\usepackage{comment}
\usepackage{fancyhdr}
\usepackage{url}
\usepackage{bm}
\usepackage{graphicx}
\usepackage{balance}
\usepackage{lastpage}

\newcommand{\tf}{TensorFlow} 
\begin{document}

\title{\bf {\tf}: A system for large-scale machine learning}
\author{Mart\'{i}n~Abadi,
Paul~Barham,
Jianmin~Chen,
Zhifeng~Chen,
Andy~Davis,
Jeffrey~Dean,
Matthieu~Devin,
Sanjay~Ghemawat,
Geoffrey~Irving,
Michael~Isard,
Manjunath~Kudlur,
Josh~Levenberg,
Rajat~Monga,
Sherry~Moore,
Derek~G.~Murray,
Benoit~Steiner,
Paul~Tucker,
Vijay~Vasudevan,
Pete~Warden,
Martin~Wicke,
Yuan~Yu, and
Xiaoqiang~Zheng\newline \newline
Google Brain
\vspace{-1mm}
}
\date{}
\maketitle
\thispagestyle{empty}

\begin{abstract}
{\tf} is a machine learning system that operates at large scale and in
heterogeneous environments.  {\tf} uses dataflow graphs to represent
computation, shared state, and the operations that mutate that
state. It maps the nodes of a dataflow graph across many machines in a
cluster, and within a machine across multiple computational devices,
including multicore CPUs, general-purpose GPUs, and custom designed
ASICs known as Tensor Processing Units (TPUs). This architecture gives
flexibility to the application developer: whereas in previous
``parameter server'' designs the management of shared state is built
into the system, {\tf} enables developers to experiment with novel
optimizations and training algorithms. {\tf} supports a variety of
applications, with particularly strong support for training and
inference on deep neural networks. Several Google services use {\tf}
in production, we have released it as an open-source project, and it
has become widely used for machine learning research. In this paper,
we describe the {\tf} dataflow model in contrast to existing systems,
and demonstrate the compelling performance that {\tf} achieves for
several real-world applications.
\end{abstract}
\vspace{-2mm}

\input{intro}
\input{bg}
\input{model}
\input{graph}
\input{impl}
\input{eval}
\input{conc}

\section*{Acknowledgments}

We gratefully acknowledge contributions from our colleagues within
Google, and from members of the wider machine learning community.  In
particular, we appreciate the feedback we have received both from the
rest of the Google Brain team and the hundreds of DistBelief and {\tf}
users that has helped us improve the usability of functionality of the
system.

Many individuals have contributed to {\tf}, including: John
Giannandrea (for creating a supportive research environment); Irina
Kofman, Amy McDonald Sandjideh, and Phing Turner (project management);
Ashish Agarwal, Dave Andersen, Anelia Angelova, Eugene Brevdo,
Yaroslav Bulatov, Jerjou Cheng, Maciek Chociej, Craig Citro, Greg
Corrado, George Dahl, Andrew Dai, Lucy Gao, mig Gerard, Ian
Goodfellow, Stephan Gouws, Gunhan Gulsoy, Steinar Gunderson, Andrew
Harp, Peter Hawkins, Yangqing Jia, Rafal Jozefowicz, \L{}ukasz Kaiser,
Naveen Kumar, Geoffrey Hinton, Mrinal Kalakrishnan, Anjuli Kannan,
Rasmus Larsen, Yutaka Leon-Suematsu, Frank Li, Peter Liu, Xiaobing
Liu, Olivia Nordquist, Chris Olah, Nishant Patil, Saurabh Saxena, Mike
Schuster, Andrew Selle, Pierre Sermanet, Noam Shazeer, Jonathon
Shlens, Jascha Sohl-Dickstein, Ilya Sutskever, Kunal Talwar, Philip
Tucker, Vincent Vanhoucke, Oriol Vinyals, Chad Whipkey, Yonghui Wu, Ke
Yang, Zongheng Yang, and Yao Zhang (general contributions to the
project); Shan Carter, Doug Fritz, Patrick Hurst, Dilip Krishnan, Dan
Man\'{e}, Daniel Smilkov, Fernanda Vi\'{e}gas, Martin Wattenberg,
James Wexler, Jimbo Wilson, Kanit Wongsuphasawat, Cassandra Xia, and
the Big Picture team (visualization); Chris Leary, Robert Hundt,
Robert Springer, Cliff Young, and the Stream Executor team
(accelerator support); Norm Jouppi and the team that created the
Tensor Processing Unit; Kayur Patel, Michael Piatek, and the coLab
team; and the growing community of open-source contributors and users
who have helped make {\tf} better.

\bibliography{tensorflow}
\bibliographystyle{abbrv}
\balance

\end{document}

%% file: intro.tex
\section{Introduction}\label{sec:intro}

In recent years, machine learning has driven advances in many different fields~\cite{angelova2015pedestrian,ba2014multiple,frome2013devise,gonzalez2015frame,deepSpeechReviewSPM2012,heigold2013multilingual,karpathy2014large,QuocLe-ICML2012,maddison2014move,Mikolov-et-al-ICLR2013,nair2015massively,Sutskever-et-al-NIPS2014,Szegedy-et-al-CVPR2015,Vinyals-et-al-Grammar-2014,Zeiler-et-al-ICASSP-2013}.
We attribute this success to the invention of more sophisticated
machine learning models~\cite{krizhevsky2012imagenet,mnih2015human},
the availability of large datasets for tackling problems in these
fields~\cite{chelba2013lm1b,ILSVRC15}, and the development of software
platforms that enable the easy use of large amounts of computational
resources for training such models on these large
datasets~\cite{chilimbi2014project,Dean-et-al-NIPS2012}.

We introduce the {\tf} system\footnote{{\tf} can be downloaded from \url{https://github.com/tensorflow/tensorflow}.} for experimenting with new models,
training them on large datasets, and moving them into production. We
have based {\tf} on years of experience with our first-generation
system, DistBelief~\cite{Dean-et-al-NIPS2012}, both simplifying and
generalizing it to enable researchers to explore a wider variety of
ideas with relative ease.  {\tf} supports both large-scale training
and inference: it efficiently uses hundreds of powerful (GPU-enabled)
servers for fast training, and it runs trained models for inference in
production on various platforms, ranging from large distributed
clusters in a datacenter, down to performing inference locally on
mobile devices.  At the same time, it is flexible and general enough
to support experimentation and research into new machine learning
models and system-level optimizations.

{\tf} uses a unified dataflow graph to represent both the computation
in an algorithm \emph{and} the state on which the algorithm operates.
We draw inspiration from the high-level programming models of dataflow
systems~\cite{alrfou2016theano,dean2004mapreduce,zaharia2012resilient},
and the low-level efficiency of \emph{parameter
  servers}~\cite{chilimbi2014project,Dean-et-al-NIPS2012,li2014parameterserver}.  Unlike traditional dataflow systems, in which
graph vertices represent functional computation on immutable data,
{\tf} allows vertices to represent computations that own or update
mutable state. Edges carry \emph{tensors} (multi-dimensional arrays)
between nodes, and {\tf} transparently inserts the
appropriate communication between distributed subcomputations.
By unifying the computation and state management in a single
programming model, {\tf} allows programmers to experiment with
different parallelization schemes that, for example, offload
computation onto the servers that hold the shared state to reduce the
amount of network traffic. We have also built various coordination
protocols, and achieved encouraging results with synchronous
replication, echoing recent
results~\cite{chen2016revisiting,cui2016geeps} that contradict the
commonly held belief that asynchronous replication is required for scalable
learning~\cite{chilimbi2014project,Dean-et-al-NIPS2012,li2014parameterserver}.

Over the past year, more than 60 teams at Google have used {\tf}, and
we have released the system as an open-source project. Thanks to our
large community of users we have gained experience with many different
machine learning applications. In this paper, we focus on neural
network training as a challenging systems problem, and select two
representative applications from this space: image classification and
language modeling. These applications stress computational throughput
and aggregate model size respectively, and we use them both to
demonstrate the extensibility of {\tf}, and to evaluate the efficiency
and scalability of our present implementation.

%% file: bg.tex
\section{Background \& Motivation}\label{sec:bg}

To make the case for developing {\tf}, we start by outlining the
requirements for a large-scale machine learning system
(\S\ref{ss:bg:reqs}), then consider how related work meets or does not
meet those requirements (\S\ref{ss:bg:rw}).

\subsection{Requirements}\label{ss:bg:reqs}

\paragraph{Distributed execution}

A cluster of powerful computers can solve many machine learning
problems more efficiently, using more data and larger models.

Machine learning algorithms generally perform better with more
training data. For example, recent breakthroughs in image
classification models have benefited from the public ImageNet dataset,
which contains 136 gigabytes of digital images~\cite{ILSVRC15}; and
language modeling has benefited from efforts like the One Billion Word
Benchmark~\cite{chelba2013lm1b}. The scale of these datasets motivates
a \emph{data-parallel} approach to training: a distributed file system
holds the data, and a set of workers processes different subsets of
data in parallel. Data-parallelism eliminates the I/O
bottleneck for input data, and any preprocessing operations can be
applied to input records independently.

Effective learned models for image recognition, language modeling,
document clustering, and many other problems have a large number of
parameters. For example, the current state-of-the-art image
classification model, ResNet, uses 2.3 million floating-point
parameters to classify images into one of 1000
categories~\cite{he2015resnet}. The One Billion Word Benchmark has a
vocabulary of 800,000 words, and it has been used to train language
models with 1.04 billion parameters~\cite{jozefowicz2016exploring}. A
distributed system can shard the model across many processes, to
increase the available network bandwidth when many workers are
simultaneously reading and updating the model.

A distributed system for model training must use the network
efficiently. Many scalable algorithms train a model using
\emph{mini-batch gradient
  descent}~\cite{Dean-et-al-NIPS2012,li2014minibatch}, where a worker
reads the current version of the model and a small batch of input
examples, calculates an update to the model that reduces a loss
function on those examples, and applies the update to the
model. Mini-batch methods are most effective when each worker uses the
most current model as a starting point, which requires a large amount
of data to be transferred to the worker with low latency.

\paragraph{Accelerator support}

Machine learning algorithms often perform expensive computations, such
as matrix multiplication and multi-dimensional convolution, which are
highly parallelizable, but have many data dependencies that require a
tightly coupled implementation. The recent availability of
general-purpose GPUs has provided a large number of cores that can
operate on fast local memory. For example, a single NVIDIA Titan X GPU
card has 6~TFLOPS peak performance~\cite{titanxspecs}. In 2012,
state-of-the-art results for different image classification tasks were
achieved using 16,000 CPU cores for three days~\cite{QuocLe-ICML2012},
and using two GPUs for six days~\cite{krizhevsky2012imagenet}. Since
then, GPU vendors have innovated in their support for machine
learning: NVIDIA's cuDNN library~\cite{chetlur2014cudnn} for GPU-based
neural network training accelerates several popular image models by
2--4$\times$ when using version R4 in place of
R2~\cite{chintala2016convnet}.

In addition to general-purpose devices, many special-purpose
accelerators for deep learning have achieved significant performance
improvements and power savings. At Google, our colleagues have built
the Tensor Processing Unit (TPU) specifically for machine learning,
and it achieves an order of magnitude improvement in
performance-per-watt compared to alternative state-of-the-art
technology~\cite{tpu-blogpost}.  The Movidius Deep Learning
Accelerator uses a low-power Myriad 2 processor with custom vector
processing units that accelerate many machine learning and computer
vision algorithms~\cite{movidius-dla}.  Ovtcharov \textit{et
  al.}\ have achieved significant performance improvements and power
savings for some convolutional models using field programmable gate
arrays (FPGAs)~\cite{msft2015catapultcnn}.  Since it is difficult to
predict the next popular architecture for executing machine learning
algorithms, we require that {\tf} uses a portable programming model
that can target a generic device abstraction, and allows its
operations to be specialized for new architectures as they emerge.

\paragraph{Training \& inference support}

In addition to training, scalable and high-performance
\emph{inference} is a requirement for using models in
production~\cite{crankshaw2015velox}. Depending on the nature of the
application, the inference may be required to produce results with
very low latency in an interactive service, or execute on a
disconnected mobile device. If the model is large, it might require
multiple servers to participate in each inference computation, and thus
require distributed computation support. Developers benefit when they
can use the same code to define a model for both training and
inference. Training and inference demand similar performance, so we
prefer a common well-optimized system for both computations. Since
inference can be computationally intensive (e.g., an image
classification model might perform 5 billion FLOPS per
image~\cite{szegedy2015rethinking}), it must be possible to accelerate
it with GPUs.

\paragraph{Extensibility}

Single-machine machine learning
frameworks~\cite{jia2014caffe,alrfou2016theano,collobert2002torch}
have extensible programming models that enable their users to advance
the state of the art with new approaches, such as adversarial
learning~\cite{goodfellow2014generative} and deep reinforcement
learning~\cite{mnih2015human}. We seek a system that provides the same
ability to experiment, and also allows users to scale up the same
code to run in production. The system must support expressive
control-flow and stateful constructs, while also satisfying our other
requirements.

\subsection{Related work}\label{ss:bg:rw}

\paragraph{Single-machine frameworks}

Many machine learning researchers carry out their work on a
single---often
GPU-equipped---computer~\cite{krizhevsky2014one,krizhevsky2012imagenet},
and many flexible single-machine frameworks have emerged to support
this scenario.  Caffe~\cite{jia2014caffe} is a high-performance
framework for training declaratively specified convolutional neural
networks that runs on multicore CPUs and
GPUs. Theano~\cite{alrfou2016theano} allows programmers to express a
model as a dataflow graph, and generates efficient compiled code for
training that model. Torch~\cite{collobert2002torch} has an imperative
programming model for scientific computation (including machine
learning) that supports fine-grained control over the order of
execution and memory utilization.

While these frameworks do not satisfy our requirement for distributed
execution, {\tf}'s programming model is close to Theano's dataflow
representation (\S\ref{sec:model}).

\paragraph{Batch dataflow systems}

Starting with MapReduce~\cite{dean2004mapreduce}, batch dataflow
systems have been applied to a large number of machine learning
algorithms~\cite{chu2007mapreduceml}, and more recent systems have
focused on increasing expressivity and
performance. DryadLINQ~\cite{yu2008dryadlinq} adds a high-level query
language that supports more sophisticated algorithms than MapReduce.
Spark~\cite{zaharia2012resilient} extends DryadLINQ with the ability
to cache previously computed datasets in memory, and is therefore
better suited to iterative machine learning algorithms (such as
$k$-means clustering and logistic regression) when the input data fit
in memory. Dandelion extends DryadLINQ to support generating code for
GPUs~\cite{rossbach2013dandelion} and
FPGAs~\cite{chung2013linqits}.

The principal limitation of a batch dataflow system is that it
requires the input data to be immutable, and all of the
subcomputations to be deterministic, so that the system can re-execute
subcomputations when machines in the cluster fail. This
feature---which is beneficial for many conventional workloads---makes
updating a machine learning model a heavy operation. For example, the
SparkNet system for training deep neural networks on Spark takes 20
seconds to broadcast weights and collect updates from five
workers~\cite{moritz2016sparknet}. As a result, these systems must
process larger batches in each model update step, which slows
convergence~\cite{byrd2012samplesize}. We show in
Subsection~\ref{ss:apps:image} that {\tf} can train larger models on
larger clusters with step times as short as 2 seconds.

While not a \emph{batch} dataflow system, Naiad~\cite{murray2013naiad}
augments a dataflow model with streaming execution, stateful vertices,
and structured timestamps (``timely dataflow'') that enable it to
handle incremental updates and iterative algorithms in the same
computation. Naiad represents iteration using cyclic dataflow graphs,
which together with mutable state make it possible to implement
algorithms that require millisecond-scale latencies for
coordination. Naiad is designed for computing on sparse, discrete
data, and does not support GPU (or any other form of) acceleration,
but we borrow aspects of timely dataflow iteration in
Subsection~\ref{ss:model:controlflow}.

\paragraph{Parameter servers}
Inspired by work on distributed key-value stores, a parameter server
architecture uses a set of servers to manage shared state that is
updated by a set of data-parallel workers. Unlike a standard key-value
store, the write operation in a parameter server is specialized for
parameter updates: it is typically an associative and commutative
\emph{combiner}, like addition-assignment (\texttt{+=}), that is
applied to the current parameter value and the incoming update to
produce a new parameter value.

Parameter servers emerged as an architecture for scalable topic
modeling~\cite{smola2010topicmodels}, and
our previous system DistBelief~\cite{Dean-et-al-NIPS2012} showed how a similar
architecture could be applied to deep neural network training. Project
Adam~\cite{chilimbi2014project} demonstrated an efficient parameter
server architecture for training convolutional neural networks, and Li
\textit{et al.}'s ``Parameter Server''~\cite{li2014parameterserver}
added innovations in consistency models, fault tolerance, and elastic
rescaling. Despite earlier skepticism that parameter servers would be
compatible with GPU acceleration~\cite{chilimbi2014project}, Cui
\textit{et al.}\ have recently shown that GeePS~\cite{cui2016geeps}, a
parameter server specialized for use with GPUs, can achieve speedups
on modest-sized clusters.

MXNet~\cite{chen2015mxnet} is a recent system that uses a parameter
server to scale training, supports GPU acceleration, and includes a
flexible programming model with interfaces for many languages. While
MXNet partially fulfills our extensibility requirements, the parameter
server is ``privileged'' code, which makes it difficult for
researchers to customize the handling of large models
(\S\ref{ss:graph:embedding}).

The parameter server architecture meets most of our requirements, and
our DistBelief~\cite{Dean-et-al-NIPS2012} uses parameter servers with
a Caffe-like model definition format~\cite{jia2014caffe} to great
effect. We found this architecture to be insufficiently extensible,
because adding a new optimization algorithm, or experimenting with an
unconventional model architecture would require our users to modify
the parameter server implementation, which uses C++ for performance.
While some of the practitioners who use that system are comfortable
with making these changes, the majority are accustomed to writing
models in high-level languages, such as Python and Lua, and the
complexity of the high-performance parameter server implementation is
a barrier to entry. With {\tf} we therefore sought a high-level
programming model that allows users to customize the code that runs in
all parts of the system (\S\ref{sec:model}).

%% file: model.tex
\begin{figure*}
  \centering
  \includegraphics{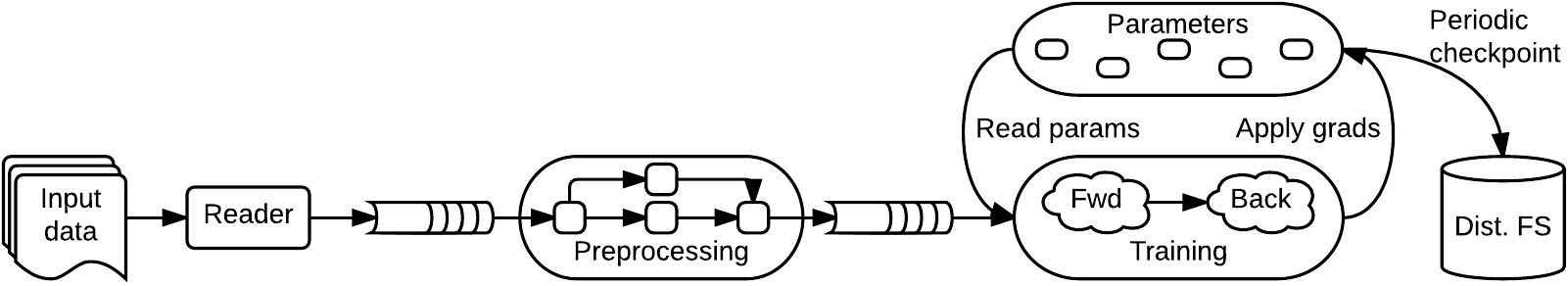}
\caption{A schematic {\tf} dataflow graph for a training pipeline
  contains subgraphs for reading input data, preprocessing, training,
  and checkpointing state.}
\label{fig:model:partial:pipeline}
\end{figure*}

\section{{\tf} execution model}\label{sec:model}

{\tf} uses a single dataflow graph to represent all computation and
state in a machine learning algorithm, including the individual
mathematical operations, the parameters and their update rules, and
the input pre-processing
(Figure~\ref{fig:model:partial:pipeline}). Dataflow makes the
communication between subcomputations explicit, and therefore makes it
easy to execute independent computations in parallel, and partition
the computation across multiple distributed devices. Dataflow {\tf}
differs from batch dataflow systems (\S\ref{ss:bg:rw}) in two respects:

\begin{itemize}
\item The model supports multiple concurrent executions on overlapping
  subgraphs of the overall graph.
\item Individual vertices may have mutable state that can be shared
  between different executions of the graph.
\end{itemize}

The key observation in the parameter server
architecture~\cite{Dean-et-al-NIPS2012,chilimbi2014project,li2014parameterserver}
is that mutable state is crucial when training very large models,
because it becomes possible to make in-place updates to very large
parameters, and propagate those updates to parallel training steps as
quickly as possible. Dataflow with mutable state enables {\tf} to
mimic the functionality of a parameter server, but with additional
flexibility, because it becomes possible to execute arbitrary dataflow
subgraphs on the machines that host the shared model parameters. As a
result, our users have been able to experiment with different
optimization algorithms, consistency schemes, and parallelization
strategies.

\subsection{Dataflow graph elements}\label{ss:model:elements}

In a {\tf} graph, each vertex represents an atomic unit of
computation, and each edge represents the output from or input to a
vertex. We refer to the computation at vertices as \emph{operations},
and the values that flow along edges as \emph{tensors}, because {\tf}
is designed for mathematical computation, and uses tensors (or
multi-dimensional arrays) to represent all data in those computations.

\paragraph{Tensors}

In {\tf}, we model all data as tensors (dense $n$-dimensional
arrays) with each element having one of a small number of primitive
types, such as \texttt{int32}, \texttt{float32}, or
\texttt{string}. Tensors naturally represent the inputs to and results
of the common mathematical operations in many machine learning
algorithms: for example, a matrix multiplication takes two 2-D tensors
and produces a 2-D tensor; and a mini-batch 2-D convolution takes two
4-D tensors and produces another 4-D tensor.

All tensors in {\tf} are dense. This decision ensures that the lowest
levels of the system can have simple implementations for memory
allocation and serialization, which reduces the overhead imposed by
the framework.  To represent sparse tensors, {\tf} offers two
alternatives: either encode the data into variable-length
\texttt{string} elements of a dense tensor, or use a tuple of dense
tensors (e.g., an $n$-D sparse tensor with $m$ non-zero elements could
be represented an $m \times n$ index matrix and a length-$m$ value
vector). The size of a tensor can vary in one or more dimensions,
making it possible to represent sparse tensors with differing numbers
of elements, at the cost of more sophisticated shape inference. %

\paragraph{Operations}

An operation takes $m \ge 0$ tensors as input, and produces $n \ge 0$
tensors as output. An operation has a named ``type'' (such as
\texttt{Const}, \texttt{MatMul}, or \texttt{Assign}) and may have zero
or more compile-time attributes that determine its behavior. An
operation can be generic and variadic at compile-time: its attributes
determine both the expected types and arity of its inputs and outputs.

For example, the simplest operation \texttt{Const} has no inputs and a
single output. \texttt{Const} has an attribute \texttt{T} that
determines the type of its output, and an attribute \texttt{Value}
that determines the value that it produces.  \texttt{AddN} is
variadic: it has a type attribute \texttt{T}, and an integer attribute
\texttt{N} that defines how many inputs (of type \texttt{T}) it
accepts.

\paragraph{Stateful operations: variables}

An operation can contain mutable state that is read and/or written
each time it executes. A \texttt{Variable} operation owns a mutable
buffer that is used to store the shared parameters of a model as it is
trained. A \texttt{Variable} has no inputs, and produces a
\emph{reference handle}, which acts as a typed capability for reading
and writing the buffer. A \texttt{Read} operation takes a reference
handle as input, and outputs the value of the variable as a dense
tensor. Several operations can modify the underlying buffer: for
example, \texttt{AssignAdd} takes a reference handle $r$ and a tensor
value $x$, and when executed performs the update
$\mathrm{State}^\prime[r] \leftarrow \mathrm{State}[r] +
x$. Subsequent \texttt{Read}$(r)$ operations produce the value
$\mathrm{State}^\prime[r]$.

\paragraph{Stateful operations: queues}

{\tf} includes several queue implementations, which support more
advanced forms of coordination. The simplest queue is
\texttt{FIFOQueue}, which owns an internal queue of tensors, and
supports concurrent access. Like a \texttt{Variable}, the
\texttt{FIFOQueue} operation produces a reference handle that can be
consumed by one of the standard queue operations, such as
\texttt{Enqueue} and \texttt{Dequeue}. These operations respectively
push their input onto the tail of the queue, or pop the head element
and output it. \texttt{Enqueue} will block if its given queue is full,
and \texttt{Dequeue} will block if its given queue is empty. When
queues are used in an input preprocessing pipeline, this blocking
provides backpressure; it also supports synchronization
(\S\ref{ss:graph:syncrep}).

\subsection{Partial and concurrent execution}\label{ss:model:partial}

{\tf} uses the dataflow graph to represent all possible computations
in a particular application, and the API for executing a graph allows
the client to specify the \emph{subgraph} that should be executed. A
subgraph is specified declaratively: the client selects zero or more
edges to \emph{feed} input tensors into the dataflow, and one or more edges
to \emph{fetch} output tensors from the dataflow; the runtime then prunes the
graph to contain the necessary set of operations. Each invocation of
the API is called a \emph{step}, and {\tf} supports multiple
\emph{concurrent steps} on the same graph, where stateful operations
enable coordination between the steps.

Figure~\ref{fig:model:partial:pipeline} shows a typical training
application, with multiple subgraphs that execute concurrently, and
interact through shared variables and queues.  The core training
subgraph depends on a set of model parameters, and input batches from
a queue.  Many concurrent steps of the training subgraph update the
model based on different input batches, to implement data-parallel
training. To fill the input queue, concurrent preprocessing steps
transform individual input records (e.g., decoding images and applying
random distortions), and a separate I/O subgraph reads records from a
distributed file system. A checkpointing subgraph runs periodically
for fault tolerance (\S\ref{ss:graph:ft}).

Partial and concurrent execution is responsible for much of {\tf}'s
flexibility. Adding mutable state and coordination via queues
makes it possible to specify a wide variety of model architectures in
``unprivileged'' code, which enables advanced users to experiment
without modifying the internals of the {\tf} runtime.

\subsection{Distributed execution}\label{ss:model:dist}

Dataflow simplifies distributed execution, because it makes
communication between subcomputations explicit. In principle, the same
{\tf} program can be deployed to a distributed cluster of GPUs for
training, a cluster of TPUs for serving, and a cellphone for mobile
inference.

Each operation resides on a particular \emph{device}, such as a CPU or
GPU in a particular \emph{task}. A device is responsible for executing
a \emph{kernel} for each operation assigned to it. {\tf} allows
multiple kernels to be registered for a single operation, with
specialized implementations for a particular device or data type (see
\S\ref{sec:impl} for details). For many operations, such as
element-wise operators (\texttt{Add}, \texttt{Sub}, etc.), we use a
single kernel implementation that can be compiled for CPU and GPU
using different compilers.

The {\tf} runtime places operations on devices, subject to implicit or
explicit device constraints in the graph.  The placement algorithm
computes a feasible set of devices for each operation, calculates the
sets of operations that must be colocated, and selects a satisfying
device for each colocation group. Stateful operations and operations
their state must be placed on the same device, which leads to implicit
colocation constraints. In addition, the user may specify partial
device preferences such as ``any device in a particular task'', or ``a
GPU in any task'', and the runtime will respect these constraints. A
typical training application will use client-side programming
constructs to add constraints such that, for example, parameters are
distributed among a set of ``PS'' tasks.

Once the operations in a graph have been placed, and the partial
subgraph has been computed for a step (\S\ref{ss:model:partial}),
{\tf} partitions the operations into per-device subgraphs. A
per-device subgraph for device $d$ contains all of the operations that
were assigned to $d$, with additional \texttt{Send} and \texttt{Recv}
operations that replace edges across device boundaries.  \texttt{Send}
transmits its single input to a specified device as soon as the tensor
is available, using a \emph{rendezvous key} to name the
value. \texttt{Recv} has a single output, and blocks until the value
for a specified rendezvous key is available locally, before producing
that value. \texttt{Send} and \texttt{Recv} have specialized
implementations for several device-type pairs; we describe some of
these in Section~\ref{sec:impl}.

We optimized {\tf} for executing large subgraphs repeatedly with low
latency. Once the graph for a step has been pruned, placed, and
partitioned, its subgraphs are cached in their respective devices. A
client \emph{session} maintains the mapping from step definitions to
cached subgraphs, so that a distributed step on a large graph can be
initiated with one small message to each participating task.  This
model favors static, reusable graphs, but it can support dynamic
computations using dynamic control flow, as the next subsection
describes.

\subsection{Dynamic control flow}\label{ss:model:controlflow}

Most evaluation in {\tf} is \emph{strict}: all inputs to an operation
must be computed before the operation executes.  Advanced
algorithms---such as efficiently training a recurrent neural
network~\cite{jordan1986serialorder}---require dynamic control flow,
which for efficiency requires non-strict evaluation.

\begin{figure}
\centering
\includegraphics{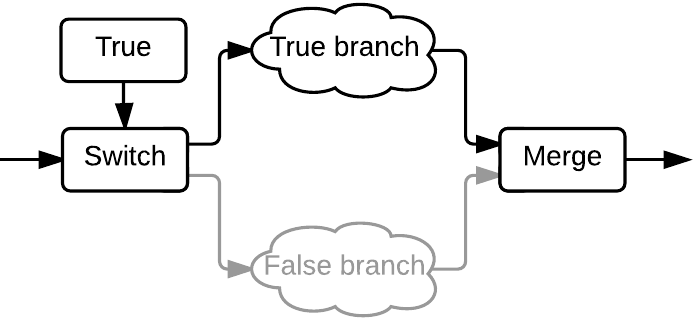}
\caption{A conditional graph using \texttt{Switch} and
  \texttt{Merge}}\label{fig:model:controlflow:cond}
\end{figure}

{\tf} supports conditional control flow using the primitive
\texttt{Switch} and \texttt{Merge} operations, which are based on
Arvind and Culler's original dynamic dataflow
architectures~\cite{Arvind1986}. \texttt{Switch} acts like a
demultiplexer: it takes a data input and a control input, and uses the
control input to select which of its two outputs should produce a
value. The \texttt{Switch} output not taken receives a special
\emph{dead} value, which propagates recursively through the rest of
the graph until it reaches a \texttt{Merge} operation. \texttt{Merge}
acts like a multiplexer: it forwards at most one non-dead input to its
output, or produces a dead output if both of its inputs are dead. We use
these primitives to build a non-strict conditional subgraph
(Figure~\ref{fig:model:controlflow:cond}) that executes one of two
branches, based on the runtime value of a tensor.

\texttt{Switch} and \texttt{Merge} also support iteration. The
implementation of loops in {\tf} is based on \texttt{Switch} and
\texttt{Merge}~\cite{Arvind1986}, with additional structural
constraints based on timely dataflow~\cite{murray2013naiad} to
simplify the distributed execution state. Like timely dataflow, {\tf}
supports multiple concurrent iterations and nested loops, but
simplifies memory management by restricting each operation to
producing a single value per output per iteration.

%% file: graph.tex
\section{Extensibility case studies}\label{sec:graph}

By choosing a unified dataflow graph to represent all computation in
{\tf}, we have enabled users to experiment with features that were
built into the runtime of our previous system~\cite{Dean-et-al-NIPS2012}.
In this section, we discuss four extensions to {\tf} that we have built
using simple dataflow primitives and ``user-level'' code.

\subsection{Differentiation and optimization}\label{ss:graph:autodiff}

Many learning algorithms train a set of parameters using some variant
of stochastic gradient descent (SGD), which entails computing the
gradients of a cost function with respect to those parameters, then
updating the parameters based on those gradients. We implement a
user-level library for {\tf} that automatically differentiates
expressions. A user can, for example, define a neural network as a
composition of layers and a loss function, and the library will derive
the backpropagation~\cite{rumelhart1988learning}.

The differentiation algorithm performs breadth-first search to
identify all of the backwards paths from the target operation (e.g., a
loss function) to a set of parameters, and sums the partial gradients
that each path contributes. Our users frequently specialize the
gradients for some operations, and they have implemented optimizations
like batch normalization~\cite{DBLP:journals/corr/IoffeS15} and
gradient clipping~\cite{pascanu2013difficulty} to accelerate training
and make it more robust. We have extended the algorithm to
differentiate conditional and iterative subcomputations
(\S\ref{ss:model:controlflow}), and developed techniques for managing
GPU memory when iterating (and accumulating intermediate values) over
long sequences in the input data (similar to
GeePS~\cite{cui2016geeps}).

{\tf} users can also experiment with a wide range of
\emph{optimization algorithms}, which compute new values for the
parameters in each training step. SGD is easy to implement in a
parameter server: for each parameter $W$, gradient $\partial L /
\partial W$, and learning rate $\alpha$, the update rule is $W^\prime
\leftarrow W - \alpha \times \partial L / \partial W$.  A parameter
server can implement SGD by using \texttt{-=} as the write operation,
and writing $\alpha \times \partial L / \partial W$ to each $W$ after
a training step.

However, there are many more advanced optimization schemes that are
difficult to express as a single write operation.  For example, the
Momentum algorithm accumulates a ``velocity'' for each parameter based
on its gradient over multiple iterations, then computes the parameter
update from that accumulation; and many refinements to this algorithm
have been proposed~\cite{sutskever2013momentum}. To implement Momentum
in DistBelief~\cite{Dean-et-al-NIPS2012}, we had to modify the
C++ code of the parameter server to change the representation of
parameter data, and execute arbitrary code in the write operation;
such modifications are beyond the majority of our users. Optimization
algorithms are the topic of active research, and our users have
implemented several on top of {\tf}, including Momentum, Adagrad,
Adadelta, RMSProp, Adam, and L-BFGS. These can be built in {\tf} using
\texttt{Variable} operations and primitive mathematical operations
without needing to modify the underlying system, which makes it easy
to experiment with new algorithms as they emerge.

\subsection{Handling very large models}\label{ss:graph:embedding}

\begin{figure}
\centering
\includegraphics{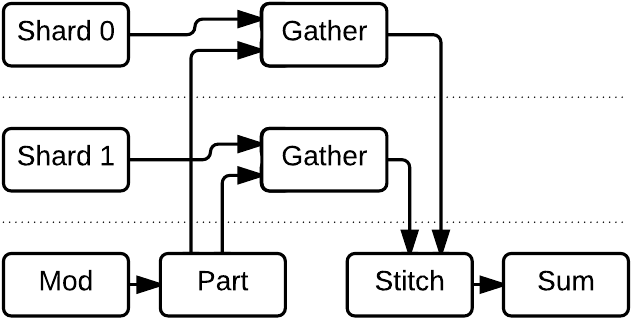}
\caption{Schematic dataflow graph for a sparse embedding layer
  containing a two-way sharded embedding
  matrix.}\label{fig:graph:embedding:lookupgraph}
\end{figure}

To train a model on high-dimensional data, such as words in a corpus
of text~\cite{brants2006ngram}, it is common to use a
\emph{distributed representation}, which embeds a training example as
a pattern of activity across several neurons, which can be learned by
backpropagation~\cite{hinton1986distributed}. For example, in a
language model, a training example might be a sparse vector with
non-zero entries corresponding to the IDs of words in a vocabulary,
and the distributed representation for each word will be a
lower-dimensional vector~\cite{bengio2003neural}.

Inference proceeds by multiplying a batch of $b$ sparse vectors
against an $n \times d$ \emph{embedding matrix}, where $n$ is the
number of words in the vocabulary, and $d$ is the desired
dimensionality, to produce a much smaller $b \times d$ dense matrix
representation; for training, most optimization algorithms modify only
the rows of the embedding matrix that were read by the sparse
multiplication. In many {\tf} models that process sparse data, $n
\times d$ can amount to gigabytes of parameters: e.g., a large
language model may use over $10^9$ parameters with a vocabulary
of 800,000 words~\cite{jozefowicz2016exploring}, and we have
experience with document models~\cite{dai2015documentembedding}
where the parameters occupy several terabytes. Such models
are too large to copy to a worker on every use, or even to store in
RAM on a single host.

We implement sparse embedding layers in the {\tf} graph as a
composition of primitive
operations. Figure~\ref{fig:graph:embedding:lookupgraph} shows a
simplified graph for an embedding layer that is split across two
parameter server tasks. The core operation of this subgraph is
\texttt{Gather}, which extracts a sparse set of rows from a tensor,
and {\tf} colocates this operation with the variable on which it
operates. The dynamic partition (\texttt{Part}) operation divides the
incoming indices into variable-sized tensors that contain the indices
destined for each shard, and the dynamic static (\texttt{Stitch})
operation reassembles the partial results from each shard into a
single result tensor. Each of these operations has a corresponding
gradient, so it supports automatic differentiation
(\S\ref{ss:graph:autodiff}), and the result is a set of sparse update
operations that act on just the values that were originally gathered
from each of the shards.

While sparse reads and updates are possible in a parameter
server~\cite{li2014parameterserver}, {\tf} adds the flexibility to
offload arbitrary computation onto the devices that host the shared
parameters. For example, classification models typically use a softmax
classifier that multiplies the final output by a weight matrix with
$c$ columns, where $c$ is the number of possible classes; for a
language model, $c$ is the size of the vocabulary, which can be
large. Our users have experimented with several schemes to accelerate
the softmax calculation. The first is similar to an optimization in
Project Adam~\cite{chilimbi2014project}, whereby the weights are
sharded across several tasks, and the multiplication and gradient
calculation are colocated with the shards. More efficient training is
possible using a \emph{sampled softmax}~\cite{jean2015nmt}, which
performs a sparse multiplication based on the true class for an
example and a set of randomly sampled false classes. We compare the
performance of these two schemes in \S\ref{ss:eval:lm1b}.

\subsection{Fault tolerance}\label{ss:graph:ft}

Training a model can take several hours or days, even using a large
number of machines~\cite{Dean-et-al-NIPS2012,chilimbi2014project}. It
is desirable to be able to train a model using non-dedicated
resources, for example using a cluster manager, like
Mesos~\cite{hindman2011mesos} or Borg~\cite{verma2015large}, that does
not guarantee availability of the same resources for the duration of
the training process. Therefore, a {\tf} job is likely to experience
failure during the training process, and we require some form of fault
tolerance.  However, failures are unlikely to be so common that
individual operations need fault tolerance, so a mechanism like
Spark's RDDs~\cite{zaharia2012resilient} would impose significant
overhead for little benefit. There is no need to make every write to
the parameter state durable, because we can recompute any update from
the input data, and many learning algorithms do not require strong
consistency~\cite{recht2011hogwild}.  Although we do not use strong
consistency for the training state, we rely on a system like
Chubby~\cite{burrows2006chubby} or ZooKeeper~\cite{hunt2010zookeeper}
to map task IDs to IP addresses.

\begin{figure*}
  \begin{center}
  \includegraphics{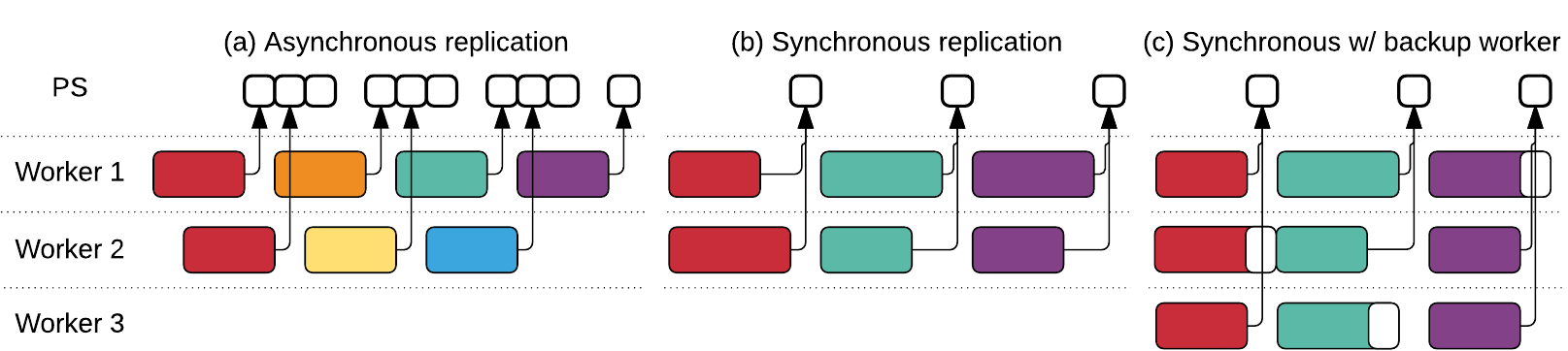}
  \end{center}
\vspace{-3mm}
\caption{Three parameter synchronization schemes for a single
  parameter in data-parallel training (\S\ref{ss:graph:syncrep}): (a)
  asynchronous, (b) synchronous without backup workers, and (c)
  synchronous with backup
  workers.}\label{fig:graph:syncrep:timelines}
\vspace{-3mm}
\end{figure*}

We implement user-level checkpointing for fault tolerance in {\tf},
using primitive operations in the graph
(Figure~\ref{fig:model:partial:pipeline}): \texttt{Save} writes one or
more tensors to a checkpoint file, and \texttt{Restore} reads one or
more tensors from a checkpoint file. Our typical configuration
connects each \texttt{Variable} in a task to the same \texttt{Save}
operation, with one \texttt{Save} per task, to maximize the I/O
bandwidth to a distributed file system. The \texttt{Restore}
operations read named tensors from a file, and a standard
\texttt{Assign} stores the restored value in its respective
variable. During training, a typical client runs all of the
\texttt{Save} operations periodically to produce a new checkpoint;
when the client starts up, it attempts to \texttt{Restore} the
latest checkpoint.

{\tf} includes a client library for constructing the appropriate graph
structure, and invoking \texttt{Save} and \texttt{Restore} as
necessary. This behavior is customizable: the user can apply different
policies to subsets of the variables in a model, or customize the
checkpoint retention scheme. For example, many users retain
checkpoints with the highest score in a custom evaluation metric. The
implementation is also reusable: it may be used for model fine-tuning
and unsupervised
pre-training~\cite{larochelle2009exploring,QuocLe-ICML2012}, which are
forms of transfer learning, in which the parameters of a model trained
on one task (e.g.\ recognizing general images) are used as the
starting point for another task (e.g.\ recognizing particular breeds
of dog).  Having checkpoint and parameter management as programmable
operations in the graph gives users the flexibility to implement
schemes like these and others that we have not anticipated.

The checkpointing library does not attempt to produce consistent
checkpoints: if training and checkpointing execute concurrently, the
checkpoint may include none, all, or some of the updates from the
training step. This is no problem for models that we train by
asynchronous gradient descent~\cite{Dean-et-al-NIPS2012}. Consistent
checkpoints require additional synchronization to ensure that
checkpointing does not run concurrently with update operations. For
example, one can use the scheme in next subsection to take a
checkpoint after the synchronous update step.

\subsection{Synchronous replica coordination}\label{ss:graph:syncrep}

SGD is robust to asynchrony~\cite{recht2011hogwild}, and previous
systems train deep neural networks using asynchronous parameter
updates~\cite{Dean-et-al-NIPS2012,chilimbi2014project}, which are
believed scalable because they maintain high throughput in the
presence of stragglers. The increased throughput comes at the cost of
training steps using stale data. Some have recently revisited the
assumption that \emph{synchronous} training does not
scale~\cite{chen2016revisiting,cui2016geeps}. Since GPUs enable
training with hundreds---rather than
thousands~\cite{QuocLe-ICML2012}---of machines, it may be possible to
train a model synchronously in less time than asynchronous training on
the same machines.

Though we designed {\tf} for asynchronous training, we have begun
experimenting with synchronous methods. The {\tf} graph enables users
to change how parameters are read and written when training a model,
and we implement three alternatives. In the asynchronous case
(Figure~\ref{fig:graph:syncrep:timelines}(a)), each worker reads the
current value when the step begins, and applies its gradient to the
different current value at the end: this ensures high utilization, but
the individual steps use stale information, making each step less
effective. The synchronous cases use queues
(\S\ref{ss:model:elements}) to coordinate execution: a blocking queue
acts as a barrier to ensure that all workers read the same parameter
version, and a second queue accumulates multiple gradient updates in
order to apply them atomically. The simple synchronous version
(Figure~\ref{fig:graph:syncrep:timelines}(b)) accumulates updates from
all workers before applying them, but slow workers limit overall
throughput.

To mitigate stragglers, we implement \emph{backup workers}
(Figure~\ref{fig:graph:syncrep:timelines}(c),~\cite{chen2016revisiting}),
which are similar to MapReduce backup
tasks~\cite{dean2004mapreduce}. Whereas MapReduce starts backup tasks
reactively---after detecting a straggler---our backup workers run
proactively, and the aggregation takes the first $m$ of $n$ updates
produced. We exploit the fact that SGD samples training data randomly,
so each worker processes a different random batch.  In
Subsection~\ref{ss:apps:image} we show how backup workers improve
throughput by up to 15\%.

%% file: impl.tex
\section{Implementation}\label{sec:impl}

\begin{figure}
\centering
\includegraphics{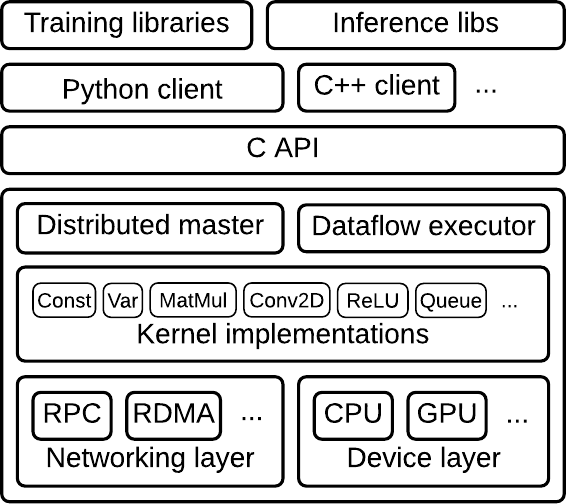}
\caption{The layered {\tf} architecture.}\label{fig:impl:0:layers}
\end{figure}

We implement {\tf} as an extensible, cross-platform
library. Figure~\ref{fig:impl:0:layers} illustrates the system
architecture: a thin C API separates user-level in various languages
from the core library. In this section, we discuss the implementation
of the various components.

The core {\tf} library is implemented in C++ for portability and
performance: it runs on several operating systems including Linux, Mac
OS X, Android, and iOS; the x86 and various ARM-based CPU
architectures; and NVIDIA's Kepler, Maxwell, and Pascal GPU
microarchitectures. The implementation is open-source, and we have
accepted several external contributions that enable {\tf} to run on
other architectures.

The \emph{distributed master} translates user requests into execution
across a set of tasks. Given a graph and a step definition, it prunes
(\S\ref{ss:model:partial}) and partitions (\S\ref{ss:model:dist}) the
graph to obtain subgraphs for each participating device, and caches
these subgraphs so that they may be re-used in subsequent steps. Since
the master sees the overall computation for a step, it applies
standard optimizations such as common subexpression elimination and
constant folding; pruning is a form of dead code elimination.
It then coordinates execution of the optimized subgraphs across a set
of tasks.

The \emph{dataflow executor} in each task handles requests from the
master, and schedules the execution of the kernels that comprise a
local subgraph. We optimize the dataflow executor for running large,
fine-grained graphs with low overhead; our current implementation
dispatches approximately 2,000,000 null operations per second. The
dataflow executor dispatches kernels to local devices and runs kernels
in parallel when possible: e.g., by using multiple cores in a CPU
device, or multiple streams on a GPU.

The runtime contains over 200 standard operations, including
mathematical, array manipulation, control flow, and state management
operations.  Many of the operation kernels are implemented using
Eigen::Tensor~\cite{eigen}, which uses C++ templates to generate
efficient parallel code for multicore CPUs and GPUs; however, we
liberally use libraries like cuDNN~\cite{chetlur2014cudnn} to
implement kernels where a more efficient specialization is
possible. We have also implemented support for \emph{quantization},
which enables faster inference in environments such as mobile devices
and high-throughput datacenter applications, and use the
\texttt{gemmlowp} low-precision matrix multiplication
library~\cite{gemmlowp} to accelerate quantized computation.

We specialize \texttt{Send} and \texttt{Recv} operations for each pair
of source and destination device types. Transfers between local CPU
and GPU devices use the \texttt{cudaMemcpyAsync()} API to overlap
computation and data transfer; transfers between two local GPUs use
DMA to relieve pressure on the host. For transfers between tasks,
{\tf} supports multiple protocols, including gRPC over TCP, and RDMA
over Converged Ethernet. We are also investigating optimizations for
GPU-to-GPU communication that use collective operations~\cite{nccl}.

Section~\ref{sec:graph} describes features that we implement totally
above the C API, in user-level code. Typically, users compose standard
operations to build higher-level abstractions, such as neural network
layers, optimization algorithms (\S\ref{ss:graph:autodiff}), and
sharded embedding computations (\S\ref{ss:graph:embedding}). {\tf}
supports multiple client languages, and we have prioritized support
for Python and C++, because our internal users are most familiar with
these languages. As features become more established, we typically
port them to C++, so that users can access an optimized implementation
from all client languages.

If it is difficult or inefficient to represent a subcomputation as a
composition of operations, users can register additional kernels that
provide an efficient implementation written in C++. We have found it
profitable to hand-implement \emph{fused kernels} for some performance
critical operations, such as a the ReLU and Sigmoid activation
functions and their corresponding gradients. We are currently
investigating automatic kernel fusion using
Halide~\cite{ragan2013halide} and other compiler-based techniques.

In addition to the core runtime, our colleagues have built several
tools that aid users of {\tf}. These include serving infrastructure
for running inference in production, a visualization dashboard that
enables users to follow the progress of a training run, a graph
visualizer that helps users to understand the connections in a model,
and a distributed profiler that traces the execution of a computation
across multiple devices and tasks. We describe these tools in an
extended whitepaper~\cite{tensorflow2015-whitepaper}, and they can be
downloaded from the project repository.

%% file: eval.tex
\section{Evaluation}\label{sec:eval}

In this section, we evaluate the performance of {\tf} on several
synthetic and realistic workloads. Unless otherwise stated, we run all
experiments on a shared production cluster, and all figures plot
median values with error bars showing the 10th and 90th percentiles.

Here we focus on system performance metrics, rather than learning
objectives like time to accuracy. {\tf} is a system that allows
machine learning practitioners and researchers to experiment with new
techniques, and this evaluation demonstrates that the system (i) has
little overhead, and (ii) can employ large amounts of computation to
accelerate real-world applications. While techniques like synchronous
replication can enable some models to converge in fewer steps overall,
we defer the analysis of such improvements to other papers.

\subsection{Single-machine benchmarks}\label{ss:eval:convnet}

Although {\tf} is a system for ``large-scale'' machine learning, it is
imperative that scalability does not mask poor performance at small
scales~\cite{mcsherry2015scalability}. Table~\ref{tbl:eval:convnet:steptimes}
contains results from Chintala's independent benchmark of
convolutional models on {\tf} and three single-machine
frameworks~\cite{chintala2016convnet}. All frameworks use a six-core
Intel Core i7-5930K CPU at 3.5\,GHz and an NVIDIA Titan X GPU.

\begin{table}[h]
\centering
{
\footnotesize
\begin{tabular}{c|rrrr}
& \multicolumn{4}{c}{Training step time (ms)} \\
\cline{2-5}
Library & AlexNet & Overfeat & OxfordNet & GoogleNet \\
\hline
Caffe~\cite{jia2014caffe} & 324 & 823 & 1068 & 1935 \\
Neon~\cite{neon} & 87 & \textbf{211} & \textbf{320} & \textbf{270} \\
Torch~\cite{collobert2002torch} & \textbf{81} & 268 & 529 & 470 \\
\hline
{\tf} & \textbf{81} & 279 & 540 & 445 \\
\end{tabular}
}
\caption{Step times for training four convolutional models with
  different libraries, using one GPU. All results are for training
  with 32-bit floats. The fastest library for each model is shown in
  bold.}\label{tbl:eval:convnet:steptimes}
\end{table}

Table~\ref{tbl:eval:convnet:steptimes} shows that {\tf} achieves
shorter step times than Caffe~\cite{jia2014caffe}, and performance
within 6\% of the latest version of
Torch~\cite{collobert2002torch}. We attribute the similar performance
of {\tf} and Torch to the fact that both use the same version of the
cuDNN library~\cite{chetlur2014cudnn}, which implements the
convolution and pooling operations on the critical path for training;
Caffe uses open-source implementations for these operations that are
simpler but less efficient than cuDNN. The Neon library~\cite{neon}
outperforms {\tf} on three of the models, by using hand-optimized
convolutional kernels~\cite{lavin2015fast} implemented in assembly
language; in principle, we could implement these kernels in {\tf}, but
we have not yet done so.

\subsection{Synchronous replica microbenchmark}\label{ss:eval:latency}

\begin{figure}
{
\centering
\includegraphics[width=\columnwidth]{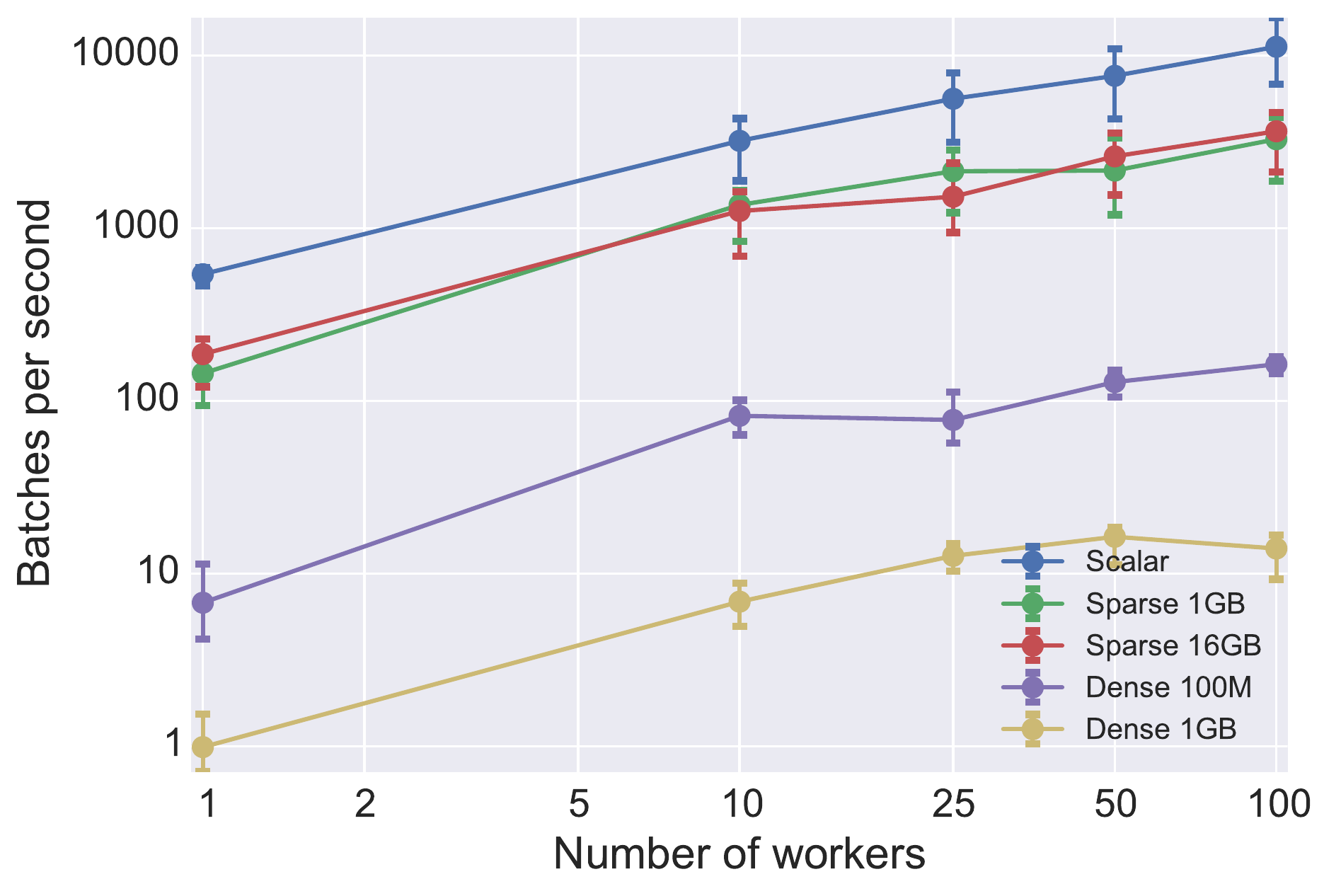}
}
\vspace{-5mm}
\caption{Baseline throughput for synchronous replication with a null
  model. Sparse accesses enable {\tf} to handle larger models, such as
  embedding matrices (\S\ref{ss:graph:embedding}).}\label{fig:eval:latency:graph}
\vspace{-2mm}
\end{figure}

The performance of our coordination implementation
(\S\ref{ss:graph:syncrep}) is the main limiting factor for scaling
with additional machines. Figure~\ref{fig:eval:latency:graph} shows
that number of \emph{null training steps} that {\tf} performs per
second for varying model sizes, and increasing numbers of
\emph{synchronous} workers. In a null training step, a worker fetches
the shared model parameters from 16 PS tasks, performs a trivial
computation, and sends updates to the parameters.

The \emph{Scalar} curve in Figure~\ref{fig:eval:latency:graph} shows
the best performance that we could expect for a synchronous training
step, because only a single 4-byte value is fetched from each PS task.
The median step time is 1.8\,ms using a single worker, growing to
8.8\,ms with 100 workers. These times measure the overhead of the
synchronization mechanism, and capture some of the noise that we
expect when running on a shared cluster.

The \emph{Dense} curves show the performance of a null step when the
worker fetches the entire model. We repeat the experiment with models
of size 100\,MB and 1\,GB, with the parameters sharded equally over 16
PS tasks.  The median step time for 100\,MB increases from 147\,ms
with one worker to 613\,ms with 100 workers. For 1\,GB, it increases
from 1.01\,s with one worker to 7.16\,s with 100 workers.

\begin{figure*}
{
\centering
\includegraphics[width=0.33\textwidth]{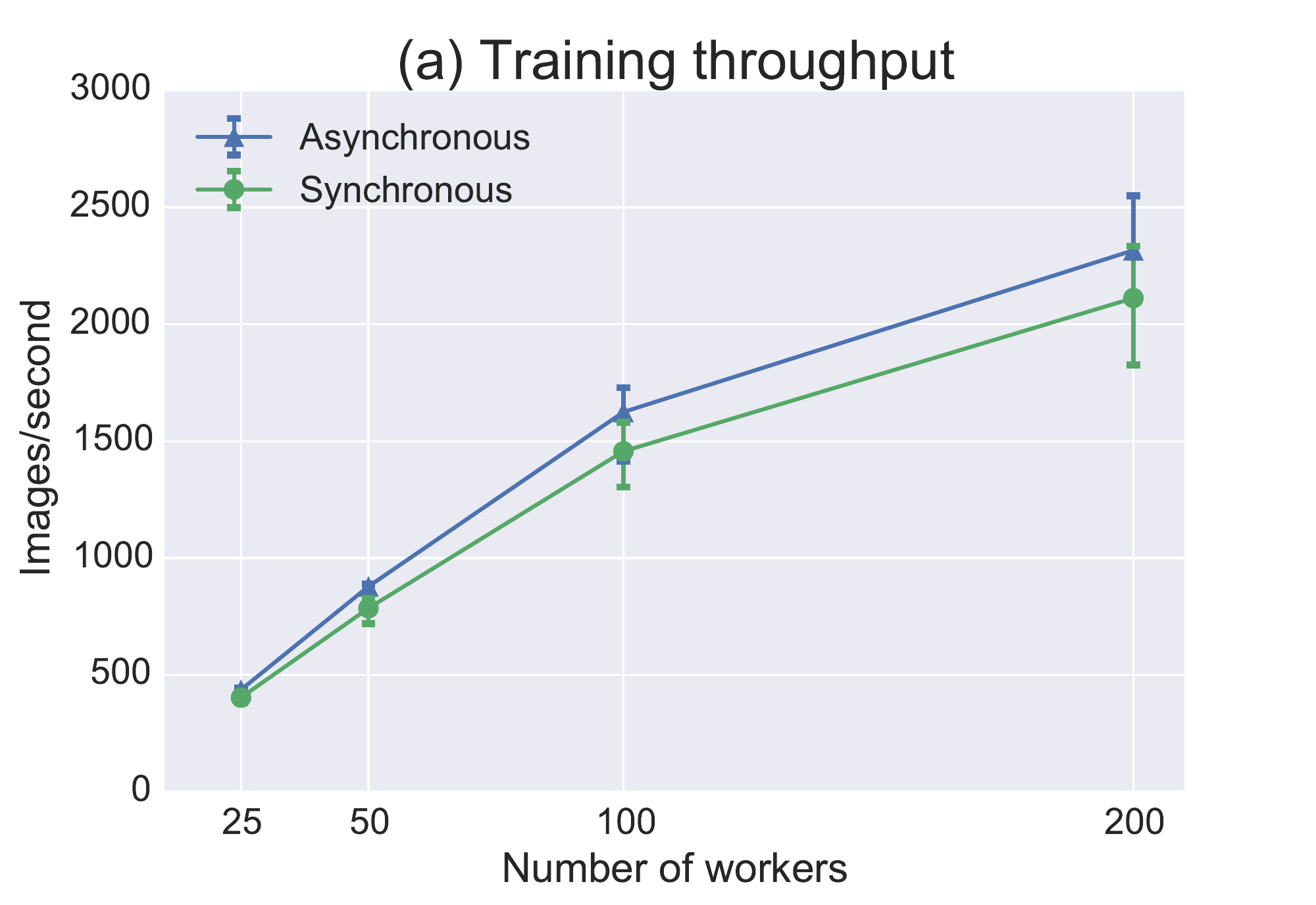}
\includegraphics[width=0.33\textwidth]{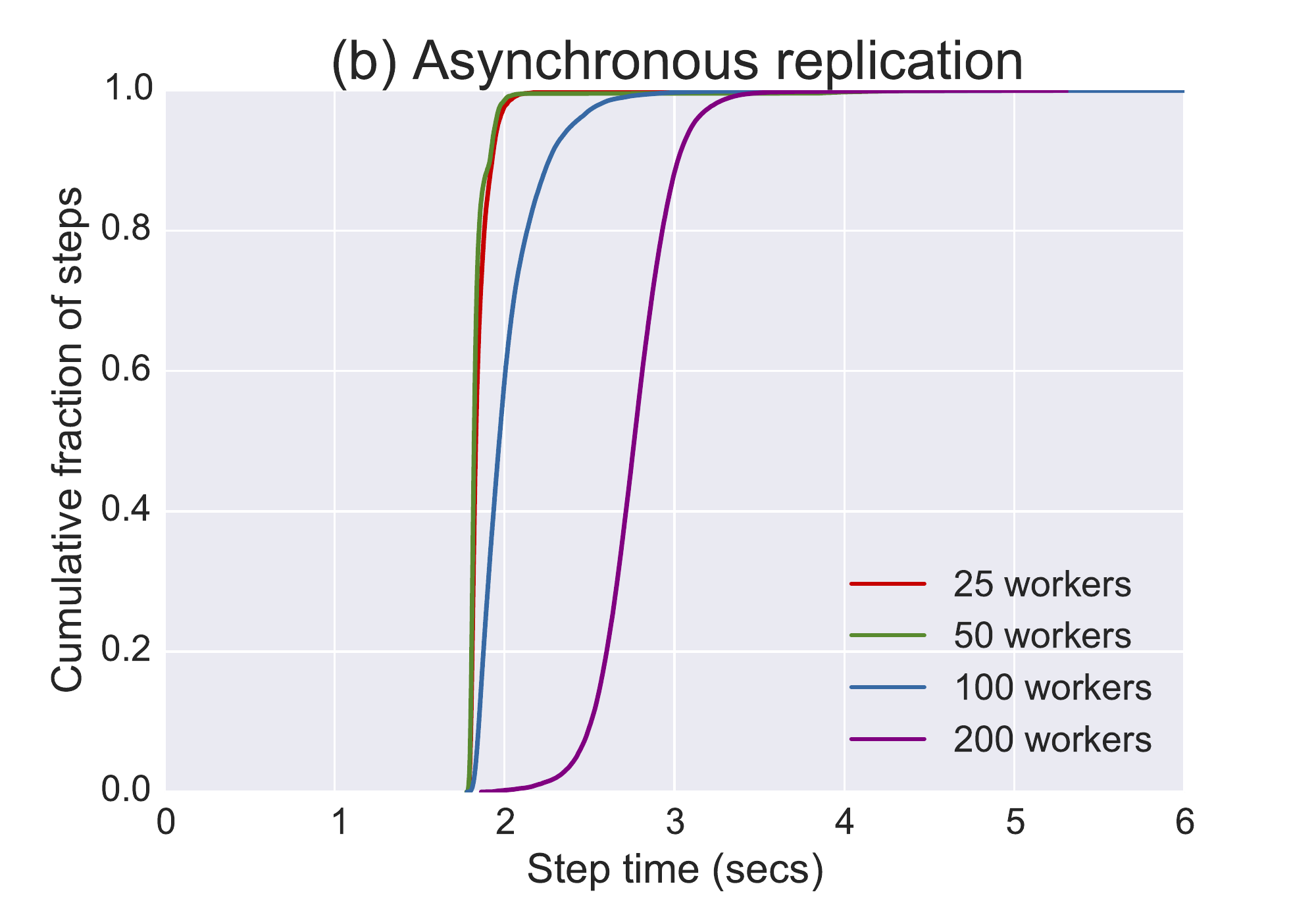}
\includegraphics[width=0.33\textwidth]{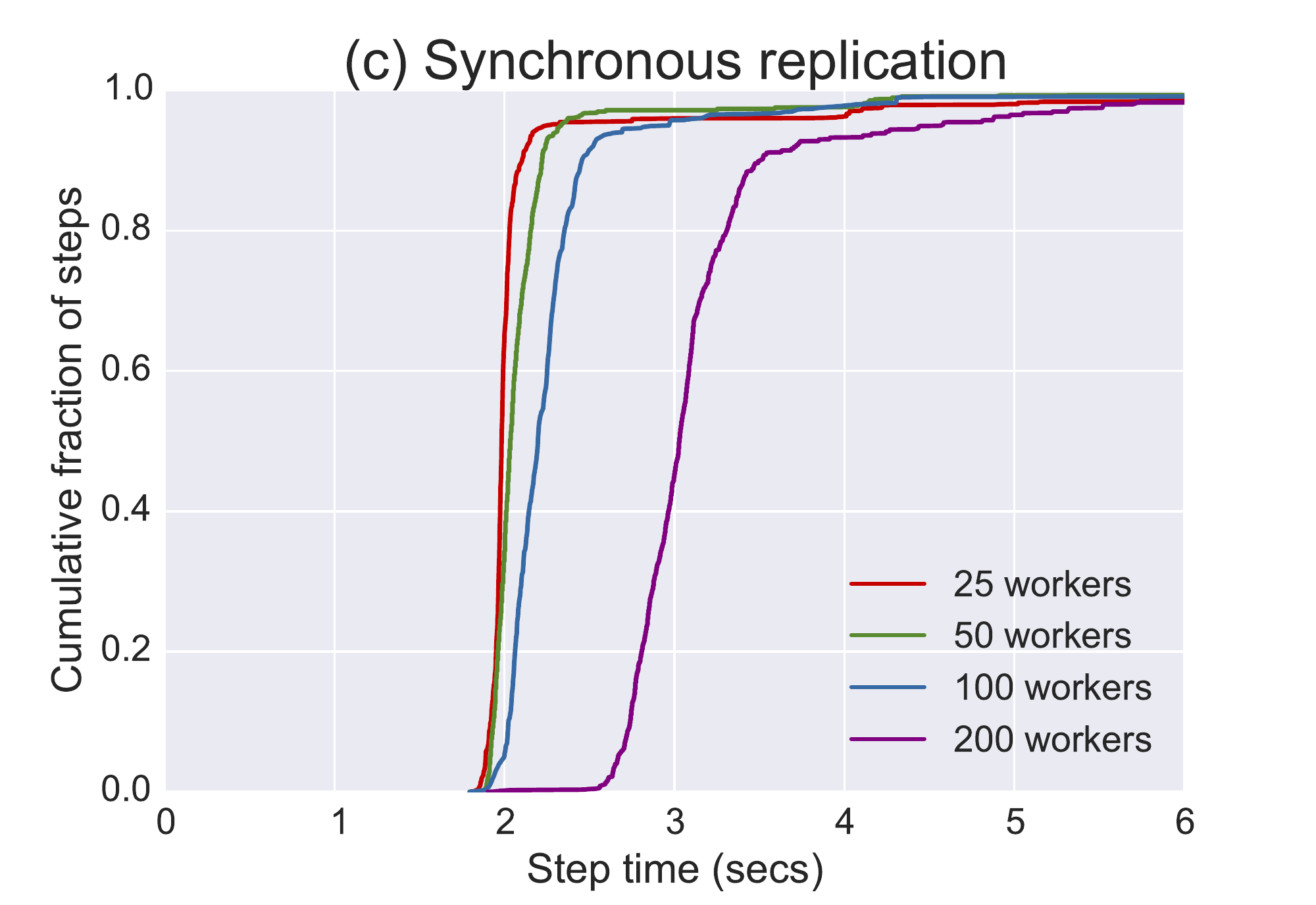}
}
\caption{(a) Inception-v3 training throughput increases with up to 200
  workers. However, adding more workers gets diminishing returns
  because the step time increases for both (b) asynchronous and (c)
  synchronous replication.}\label{fig:apps:image:all}
\end{figure*}

For large models, it is typical that a training step accesses only a
subset of the parameters, and the \emph{Sparse} curves show the
throughput of the embedding lookup operation from
Subsection~\ref{ss:graph:embedding}. Each worker reads 32 randomly
selected entries from a large embedding matrix containing 1\,GB or
16\,GB of data. As expected, the step times do not vary with the size
of the embedding, and {\tf} achieves step times ranging from 5 to
20\,ms.

\subsection{Image classification}\label{ss:apps:image}

Deep neural networks have achieved breakthrough performance on
computer vision tasks such as recognizing objects in
photographs~\cite{krizhevsky2012imagenet}, and these tasks are a key
application for {\tf} at Google.  Training a network to high
accuracy requires a large amount of computation, and we use {\tf} to
scale out the computation across a cluster of GPU-enabled servers.  In
these experiments, we focus on Google's Inception-v3 model, which
achieves 78.8\% accuracy the ILSVRC 2012 image classification
challenge~\cite{szegedy2015rethinking}; the same techniques apply to
other deep convolutional models---such as Microsoft's
ResNet~\cite{he2015resnet}---that {\tf} users have implemented.  We
investigate the scalability of training the Inception-v3 model using
multiple replicas. We configure a {\tf} job with 17 PS tasks, and vary
the number of worker tasks. Each worker task has one NVIDIA K40 GPU
and 5 IvyBridge cores, and a PS task has 8 IvyBridge cores. We investigate the
effect of coordination (\S\ref{ss:graph:syncrep}) on training
performance, using up to 200 workers to validate recent promising
results for synchronous
training~\cite{chen2016revisiting,cui2016geeps}. In particular, if
synchronous training can be made efficient, a model such as
Inception-V3 will train in fewer steps, and converge to a higher
accuracy than with asynchronous training~\cite{chen2016revisiting}.

Training throughput improves to 2,300 images per second as we increase
the number of workers to 200, but with diminishing returns
(Figure~\ref{fig:apps:image:all}(a)).
Figures~\ref{fig:apps:image:all}(b) and (c) explain the limits to
scaling: as we add more workers, the step time increases, because
there is more contention on the PS tasks, both at the network
interface and in the aggregation of updates.  As expected, for all
configurations, synchronous steps are longer than asynchronous steps,
because all workers must wait for the slowest worker to catch up
before starting the next step. While the median synchronous step is
approximately 10\% longer than an asynchronous step with the same
workers, above the 90th percentile the synchronous performance
degrades sharply, because stragglers disproportionately impact the
tail.

\begin{figure}
\begin{center}
\includegraphics[height=2in]{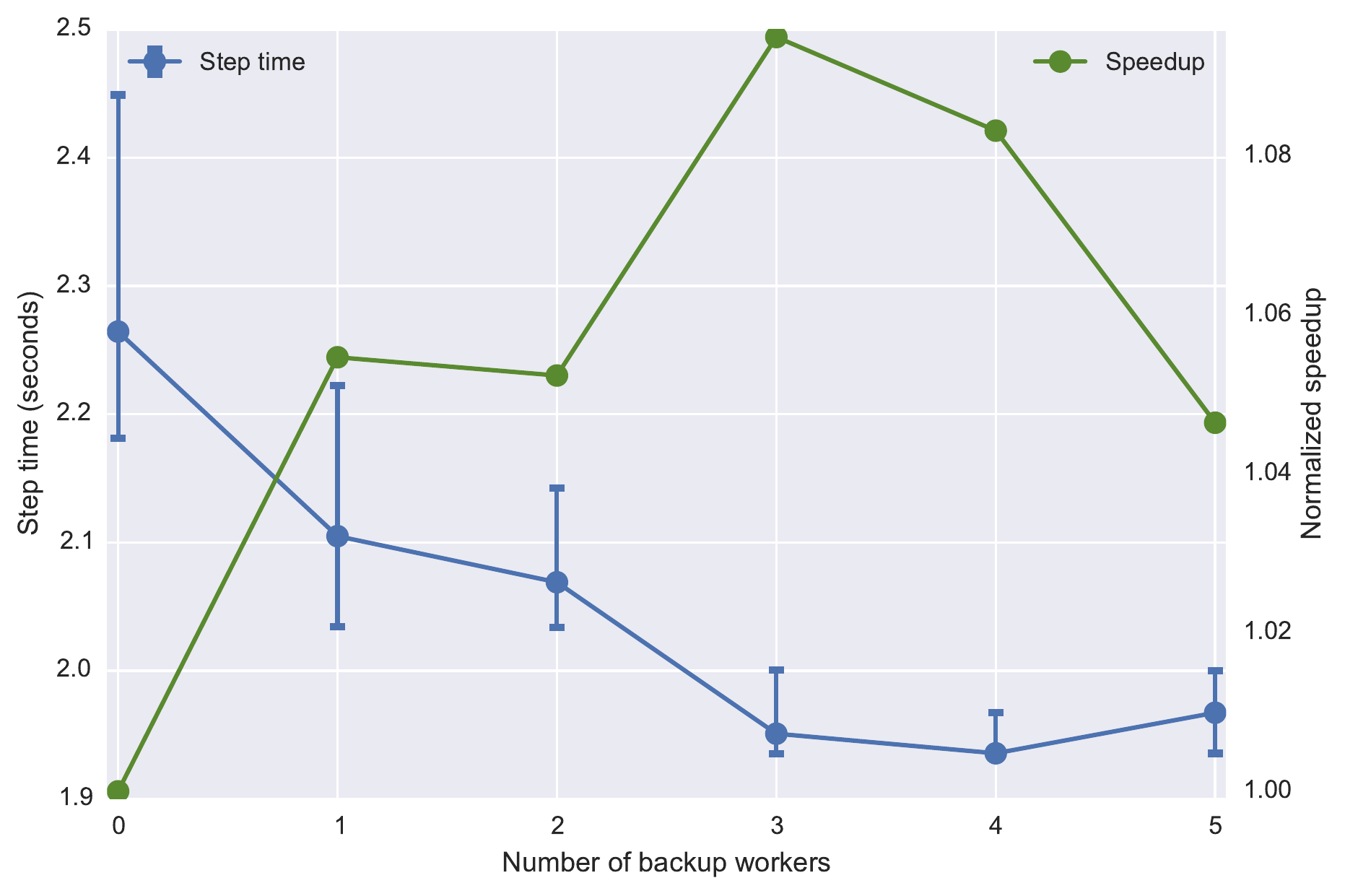}
\end{center}
\vspace{-2mm}
\caption{Backup workers reduce the step time for 50-worker
  Inception-v3 training. 4 backup workers give the shortest
  overall step time, but 3 backup workers are most efficient when
  we normalize for the total resources used.}\label{fig:apps:image:backup}
\end{figure}

To mitigate tail latency, we can add backup workers, so that a step
completes when the first $m$ of $n$ tasks produce
gradients. Figure~\ref{fig:apps:image:backup} shows the effect on step
time of adding backup workers to a 50-worker Inception training
job. Each additional backup worker up to and including the fourth
reduces the median step time, because the probability of a straggler
affecting the step decreases. Adding a fifth backup worker slightly
degrades performance, because the 51st worker (i.e., the first whose
result is discarded) is more likely to be a non-straggler that
generates more incoming traffic for the PS
tasks. Figure~\ref{fig:apps:image:backup} also plots the
\emph{normalized speedup} for each configuration, which we define as
$t(b) / t(0) \times 50 / (50 + b)$ (where $t(b)$ is the median step
time with $b$ backup workers), and which discounts the speedup by the
fraction of additional resources consumed. Although adding 4 backup
workers achieves the shortest overall step time (1.93\,s), adding 3
achieves the highest normalized speedup (9.5\%), and hence trains the
model to the same quality using less aggregate GPU-time.

\subsection{Language modeling}\label{ss:eval:lm1b}

\begin{figure}
\begin{center}
\includegraphics[height=2in]{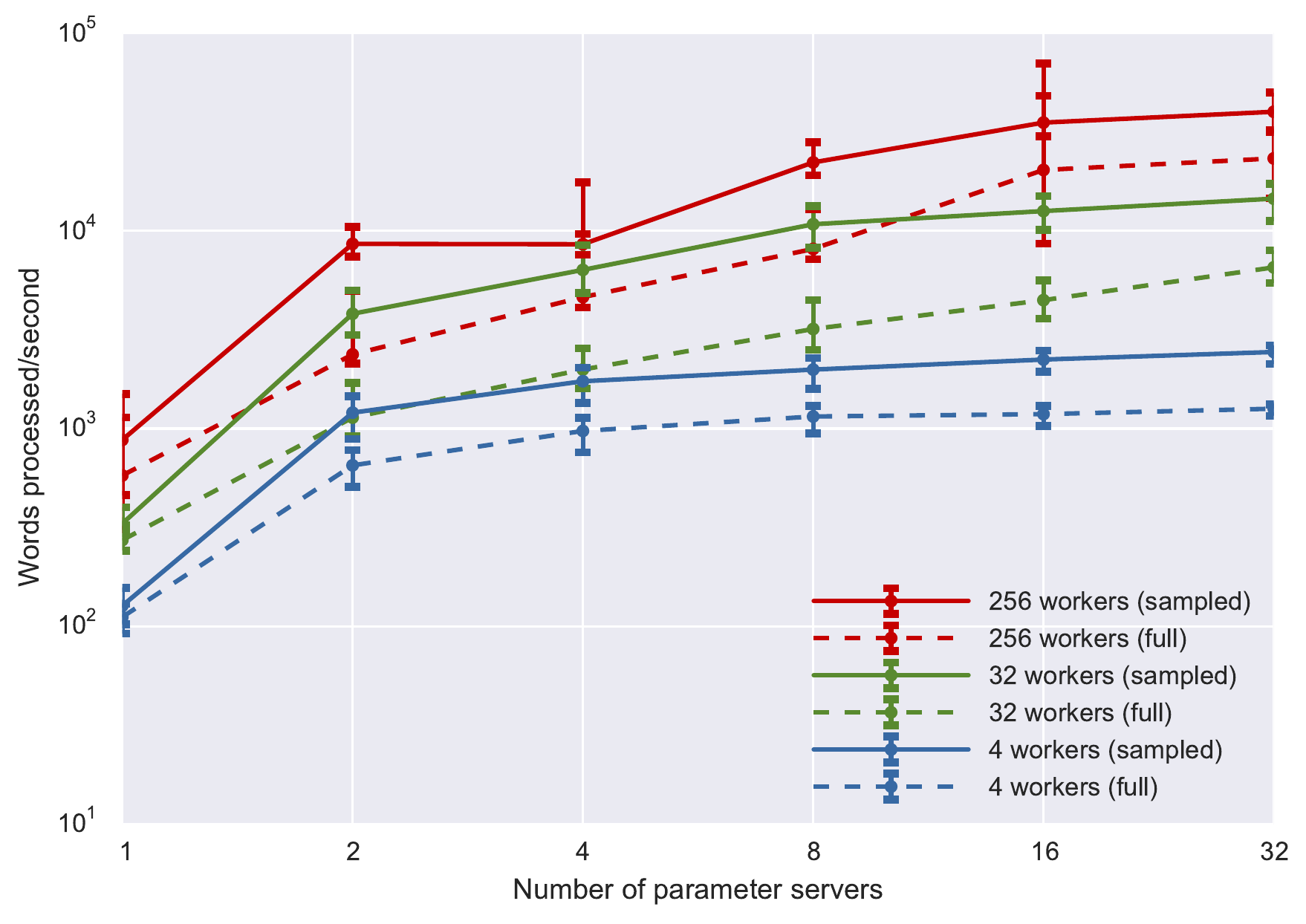}
\end{center}
\vspace{-2mm}
\caption{Increasing the number of PS tasks leads to increased
  throughput for language model training, by parallelizing the softmax
  computation. Sampled softmax increases throughput by performing less
  computation.}\label{fig:eval:lm1b:wps}
\end{figure}

Given a sequence of words, a language model predicts the most probable
next word~\cite{bengio2003neural}. Therefore, language models are
integral to predictive text, speech recognition, and translation
applications. In this experiment, we investigate how {\tf} can train a
recurrent neural network
(viz.~LSTM-512-512~\cite{jozefowicz2016exploring}) to model
the text in the One Billion Word Benchmark~\cite{chelba2013lm1b}. The
vocabulary size $|V|$ limits the performance of training, because the
final layer must decode the output state into probabilities for each
of $|V|$ classes~\cite{jean2015nmt}. The resulting parameters can be
large ($|V| \times d$ for output state dimension $d$) so we use the
techniques for handling large models from
Subsection~\ref{ss:graph:embedding}. We use a restricted vocabulary of
the most common 40,000 words---instead of the full 800,000
words~\cite{chelba2013lm1b}---in order to experiment with smaller
configurations.

Figure~\ref{fig:eval:lm1b:wps} shows the training throughput, measured
in words per second, for varying numbers of PS and worker tasks, and
two softmax implementations. The \emph{full} softmax (dashed lines)
multiplies each output by a $512 \times 40,000$ weight matrix sharded
across the PS tasks. Adding more PS tasks increases the throughput,
because {\tf} can exploit distributed model
parallelism~\cite{Dean-et-al-NIPS2012,krizhevsky2014one} and perform
the multiplication and gradient calculation on the PS tasks, as in
Project Adam~\cite{chilimbi2014project}. Adding a second PS task is
more effective than increasing from 4 to 32, or 32 to 256 workers.
Eventually the throughput saturates, as the LSTM calculations dominate
the training step.

The \emph{sampled} softmax (solid lines) reduces the data transferred
and the computation performed at the PS
tasks~\cite{jean2015nmt}. Instead of a dense weight matrix, it
multiplies the output by a random sparse matrix containing weights for
the true class and a random sample of false classes. We sample 512
classes for each batch, which reduces the softmax data transfer and
computation by a factor of 78.

%% file: conc.tex
\section{Conclusions}\label{sec:conc}

We have described the {\tf} system and its extensible dataflow-based
programming model. The core idea of this paper is that {\tf}'s
dataflow representation subsumes existing work on parameter server
systems, and offers a uniform programming model that allows users to
harness large-scale heterogeneous systems, both for production tasks
and for experimenting with new approaches. We have shown several
examples of how the {\tf} programming model supports experimentation
(\S\ref{sec:graph}) and demonstrated that the resulting
implementations are performant and scalable (\S\ref{sec:eval}).

Our initial experience with {\tf} is encouraging. A large number of
groups at Google have deployed {\tf} in production, and {\tf}
is helping our research colleagues to make new advances in machine
learning. Since we released {\tf} as open-source software, over 8,000
people have forked the source code repository, the binary distribution
has been downloaded 500,000 times, and our users have published dozens
of machine learning models that use {\tf}.

{\tf} is a work in progress. Its flexible dataflow representation
enables power users to achieve excellent performance, but we have not
yet determined default policies that work well for most users.
Further research on automatic optimization should bridge this gap.  On
the system level, we are actively developing algorithms for automatic
placement, kernel fusion, memory management, and scheduling. While the
current implementations of mutable state and fault tolerance suffice
for applications with weak consistency requirements, we expect that
some {\tf} applications will require stronger consistency, and we are
investigating how to build such policies at user-level. Finally, our
users are demanding, and some have begun to chafe at the limitations
of a static dataflow graph, especially for algorithms like deep
reinforcement learning~\cite{mnih2015human}.  Therefore, we face the
intriguing problem of providing a system that transparently and
efficiently uses distributed resources, even when the structure of the
computation unfolds dynamically.

By sharing the implementation of {\tf} and engaging with the research
community, we hope that this work will spur further research in
distributed systems and machine learning.